# Cell intrinsic dynamics guide neuroblast ingression independent of tissue fluidity

David Green[1,2], Ermia Etemadi[3], Lauren Enright[3], Rodrigo Fernandez-Gonzalez[1,4,5,6], Ulrich Tepass[1,*] and Gonca Erdemci-Tandogan[3,7,*]

[1] Department of Cell and Systems Biology, University of Toronto, Toronto, ON, M5S 3G5, Canada.

[2] Institut Pasteur, Université Paris Cité, 4D Unit, Paris, France.

[3] Department of Physics and Astronomy, University of Western Ontario, London, ON, N6A 3K7, Canada.

[4] Translational Biology and Engineering Program, Ted Rogers Centre for Heart Research, University of Toronto, Toronto, ON, M5G 1M1, Canada.

[5] Institute of Biomedical Engineering, University of Toronto, Toronto, ON, M5S 3G9, Canada.

[6] Developmental and Stem Cell Biology Program, The Hospital for Sick Children, Toronto, ON, M5G 1X8, Canada.

[7] Department of Medical Biophysics, University of Western Ontario, London, ON, N6A 3K7, Canada.

*Correspondence:
email: gerdemci@uwo.ca
email: u.tepass@utoronto.ca

## Abstract

Morphogenesis involves the coordination of multiple cellular processes that occur simultaneously within developing tissues. During early *Drosophila* embryogenesis, neuroblast (NB) ingression occurs concurrently with germ band extension (GBE), yet whether these processes interact mechanistically remains unclear. Here, we combine mathematical modelling with quantitative live imaging to investigate whether tissue-level mechanics during GBE influence NB ingression dynamics. Mathematical modelling predicted that reducing tissue fluidity through impaired cellular rearrangements should slow NB ingression by increasing mechanical resistance. Experimental analysis of mutants in which cell intercalation and GBE are disrupted revealed a dramatic reduction in tissue fluidity. However, NB ingression rates remained largely unaffected when tissue fluidity decreased. Incorporating cell-intrinsic myosin anisotropy and endocytosis-contractility coupling into our mathematical model rescued the rate of neuroblast ingression in solid-like tissues. Thus, our findings suggest that cell-intrinsic mechanisms, rather than tissue-level fluidity, maintain ingression kinetics. More broadly, these results illustrate how developmental systems can achieve robustness by insulating critical cellular events from tissue-level mechanical variability.

## Statement of Significance

Morphogenesis requires coordination of cellular processes in tissues undergoing mechanical transitions. While recent work highlights the importance of tissue fluidity in morphogenesis, whether tissue-level mechanical changes influence concurrent cellular events remains unclear. Combining mathematical modelling with quantitative live imaging, we tested whether tissue fluidization during *Drosophila* germ-band extension regulates neuroblast ingression. Shape-based vertex models failed to predict tissue mechanical states when myosin was disrupted. Instead, incorporating cellular rearrangement delays recapitulated phenotypes independent of cell shape. Strikingly, we showed that neuroblast ingression remains robust despite tissue solidification *in vivo*. Biophysical modelling suggests that coupled cell contractility and endocytosis maintains robust ingression despite tissue solidification. Our work shows that developmental programs can be mechanically insulated from tissue-scale changes, thus enabling developmental robustness.

## Introduction

Morphogenesis requires the execution of dynamic and highly coordinated cellular and tissue behaviours that sculpt tissues and organs within the developing embryo (1–4). Tissue deformation in morphogenetic processes arises from forces generated both intrinsically within cells and extrinsically by neighbouring tissues (5, 6). Tissue fluidity is increasingly recognized as a critical factor influencing morphogenesis (7–14). While individual morphogenetic events have been studied extensively in isolation, multiple such events occur simultaneously during development, raising the question of how they influence one another (15–18). Tissue deformation in one domain could mechanically influence neighboring, concurrent processes, yet what conditions determine whether these interactions occur remains poorly understood.

In early *Drosophila* embryos two temporally and spatially overlapping morphogenetic processes have been described: The ingression of neural stem cells (neuroblasts, NBs) which arise from the germband and germband extension (GBE) which elongates the anterior-posterior body axis. Here we ask whether the cell rearrangements during GBE impact NB ingression dynamics. During GBE, the epithelial germband tissue more than doubles in length along the anterior-posterior axis via convergent extension (19, 20). The major driving force of GBE is the planar polarized activity of the molecular motor non-muscle myosin II (myoII), which generates anisotropic tension across the tissue and directional cell rearrangements (21–23). In addition to intrinsic contractile forces, external inputs from the mesoderm, ectoderm and endoderm further modulate cell shapes (15, 24–27). As elongation proceeds, the germband transitions from a solid-like to fluid-like mechanical state, enabling cell intercalation and tissue remodelling (10, 11). Such transitions occur during embryonic development across various species, including insects and vertebrates (10, 12–14, 28, 29) and pathologically during cancer progression (30, 31). Here we ask whether this changing mechanical environment influences NB ingression.

Concurrent with GBE, the ventral neuroectoderm which is part of the germband produces a fixed number of NBs that delaminate in five successive waves, S1-S5 (32, 33). Here we investigate the delamination of the first wave (S1) that occurs when the germband has undergone its solid to fluid transition during extension. When NBs are specified, they begin to apically constrict, reducing their apical domain that is eventually lost as NBs delaminate basally from the epithelium (34, 35).

NB apical constriction is driven by periodic pulses of junctional and medial myosin, producing strong cortical contractions (35). Myosin dynamics in NBs are spatially organized with preferential enrichment along anterior-posterior (AP) edges (35). Remarkably, NB ingression dynamics appears normal even with substantial reduction in *myoII* activity (35), pointing to alternative mechanisms of apical domain loss. Indeed, endocytosis of the apical domain (including components of the apical membrane and apical adherens junctions) is essential for NB ingression (33). However, the interplay between actomyosin contractions and endocytic activity remain largely unclear. NB ingression speed is increased upon reducing local tissue resistance in neighboring cells, suggesting that NBs are potentially responsive to changes in tissue resistance (35). This raises the question of how NBs maintain consistent ingression dynamics in a tissue undergoing massive mechanical remodelling during GBE, whether through redundant pathways that sense and respond to tissue-level forces, or, alternatively, through cell-intrinsic mechanisms that operate independently of the surrounding tissue mechanical state. Understanding which mechanism underlies NB ingression robustness in light of dramatic changes in tissue mechanics has implications for how developmental systems insulate crucial cellular events from mechanical variability.

## Results

### Mathematical modelling predicts that tissue stiffening impairs NB ingression

To investigate the potential mechanical coupling between convergent extension and NB ingression, we developed an anisotropic vertex model that incorporates both processes (Fig. 1A, MovieS1, Methods): an anisotropic line tension reflects planar-polarized myosin distribution during germband extension, concluding in a cell intercalation; and individual cell apical area constriction through progressive area reduction initiates NB delamination. Many NBs also display the same planar-polarized organization as surrounding germband cells (35).

We first modeled tissues undergoing convergent extension with varying T1 cellular rearrangement time delay values to simulate disruptions in cell intercalation (Fig. 1A-D, MovieS1, Methods). T1 delays impose a finite cellular rearrangement time that acts as a barrier on cell rearrangements, thereby reducing tissue fluidity (36). Tissues had embedded ingressing cells with progressive area reduction to simulate NB apical constriction (Fig. 1C, MovieS1, Methods). In our model, NB constriction time increased with decreasing tissue fluidity (increasing T1 cellular rearrangement time) (Fig. 1D), suggesting that reducing tissue fluidity should create greater mechanical resistance to the cell shape changes required for NB ingression, thereby slowing delamination and potentially leading to ingression delays.

Cell shapes influence the rigidity of isotropic tissues (37). To isolate fluidity effects, we performed additional simulations using isotropic tissues in which fluidity was regulated solely through systematic variation of cell shape. These control simulations eliminated global anisotropic line tensions and directional tissue flows while maintaining other mechanical parameters. The solid-to-fluid transition in isotropic vertex model simulations is characterized by a shape parameter $p_0$ (cell shape index, cellular perimeter divided by square root of area (38)). Low $p_0$ values, corresponding to reduced cell contact, produce solid-like behavior (11). Above a critical threshold $p_0^*$, tissues become fluid-like with elongated cell configurations that facilitate neighbour exchanges (11). The isotropic simulations demonstrated that tissue fluidity can significantly impact NB ingression dynamics (Fig. 1E). NB ingression times decreased as tissues became more fluid-like (increasing cell shape index) (Fig. 1E). Simultaneous changes in the cellular rearrangement time delay and the cell shape index revealed an additive effect of the two parameters on NB ingression times (Fig. S1A-B). Taken together, the model predicts that reduced tissue fluidity, caused by impaired T1 transitions or changes in cell shape, slows NB ingression, potentially by increasing mechanical resistance (Fig. 1D–E).

**NB ingression coincides with tissue fluidization *in vivo***

To test our model predictions and explore how tissues undergo multiple simultaneous morphogenetic events, we quantified GBE, a morphogenetic event in which the *Drosophila* body axis is elongated via oriented cell-cell rearrangements (Fig. 2A,A’). We quantified GBE dynamics using live confocal imaging (Fig. 2B). We used a membrane marker (PH::mCherry) to track cell shapes and cell rearrangements in the ventral neuroectoderm during the first wave of NB ingression. Previous work showed that during early GBE, the ventral epithelium undergoes a solid-to-fluid transition as a result of cell-cell rearrangements that drive convergent extension along the anterior-posterior axis (11).

The onset of GBE is observed when the ventral epithelium begins to elongate (time = 0 min; Fig. 2C) (21). We used GBE onset to align all embryos developmentally and began analysis 2.5 min prior to GBE onset. To characterize tissue mechanics, we used three established metrics of tissue fluidity (Fig. 2D-F): instantaneous rate of cell rearrangements, cell shape index, and cell shape alignment (11). The cell shape alignment quantifies both the degree to which individual cells are elongated and how coherently those cells point in the same direction across the tissue (Methods).

Consistent with previous findings, we observed a solid-fluid transition during early GBE (Fig. 2G) (11). Cell rearrangement rates increased rapidly after GBE onset (t=0), peaking at 5-10 minutes before gradually declining (Fig. 2D). This transition coincided with progressive cell elongation (increased shape index) and reduced cell shape alignment (Fig. 2E-F).

When do NB ingressions start relative to this solid-fluid transition? Using the same time-series images analyzed above, we tracked individual NBs which were identified by their ingressive behavior (Fig. 2H). Ingression onset was defined as the time when the mean NB area began persistently declining. There was no NB ingression prior to the onset of GBE (time = 0). NBs initiated their ingression at a relatively regular pace almost immediately after the onset of GBE (Fig. 2I, MovieS2), during tissue fluidization. These findings suggest that the ingression of S1 NBs predominantly occurs in a tissue "fluidized" as a consequence of the mechanical changes associated with GBE.

**Cell morphology is decoupled from rearrangement dynamics in Myosin-depleted embryos**

To experimentally test whether tissue fluidity affects NB ingression, we sought to disrupt the solid-fluid transition during GBE. During GBE, cell-cell rearrangements are largely driven by planar polarized actomyosin cables directing contractility along the dorsoventral (DV) axis (23, 39). We first analyzed *Kruppel (Kr),* a gap gene involved in AP patterning (40). *Kr* mutant embryos do not extend their germ band fully, due to disruption in the planar polarization of myosin in the central region of the embryo (22). While the impact of *Kr* on GBE has been studied, its effect on tissue fluidity has not been reported. We imaged the ventral epithelium (Fig. 3A, MovieS3) in *Kr* mutants using PH::mCherry to track cell shapes, alignment, and rearrangement rates.

*Kr* mutants showed delayed tissue extension after the first 10 minutes of GBE (Fig. 3B,B"), consistent with previous work showing that early GBE is driven by cell shape changes while cell rearrangement becomes critical during "late" (+10 min) GBE (27, 41). Cell rearrangement rates decreased from a mean of 0.06±0.02 (per min, per cell) in control (*Kr*/+ or +/+ siblings) to 0.03±0.001 (per min, per cell) in *Kr* mutants*,* a 50% reduction (Fig. 3C,C'). *Kr* mutants showed reduced changes in cell shape index, but relatively normal shape alignment (Fig. 3D,E).

To characterize myosin changes, we tracked the cell membrane with Gap43::mCherry and imaged myosin levels using a myosin light chain (Spaghetti squash, Sqh) fusion with GFP (Sqh::GFP) (Fig. 3F). In the wild type, myosin was enriched along edges between anterior and posterior neighbors (AP edges), consistent with myosin planar polarization during GBE (Fig 3G,(21–23, 42). We found that *Kr* mutants displayed a significant reduction in the AP polarization of myosin (1.2-fold increase between AP and DV edges in wild type versus 1.06 in *Kr*) as well as a 45% reduction in total myosin (Fig 3G). To assess tissue mechanical state, we plotted the cell shape-based prediction curve for the solid-fluid transition (Fig. 3H). Wild-type embryos showed the expected close linkage between cell shape/anisotropy and rearrangement rate (Fig. 2G). However, *Kr* revealed a distinct breakdown of this relationship, with cell rearrangement rates being significantly reduced even under normal cell shape/anisotropy conditions (Fig. 3H). Together, these results indicate that *Kr* mutants show reduced tissue fluidity and that it is not fully captured by cell-shape based predictions alone.

To further enhance tissue solidification  we investigated embryos maternally depleted for myosin heavy chain, encoded by *zipper* (*zip*) using RNAi (35). Imaging the ventral epithelium with the membrane marker Spider::GFP (Fig. 4A, MovieS3), we confirmed that *zip*-RNAi caused a complete loss of posterior tissue flow and tissue elongation (Fig. 4B,B'). Cell rearrangement rates decreased dramatically from 0.07±0.01 (per min, per cell) in control embryos expressing RNAi against the *white* gene (*w*-RNAi), to 0.007±0.004 (per min, per cell) in *zip*-RNAi, a 91% reduction (Fig. 4C,C'). While rearrangement rates were severely impacted, cells still underwent relatively normal shape changes over time (Fig. 4D,E).

We next characterized the myosin dynamics in *zip*-RNAi embryos. We tracked cell membranes with Spider::GFP, and we imaged myosin with Sqh:mCherry (Fig. 4F). We found that *zip*-RNAi embryos displayed a minor change in AP myosin enrichment (1.17 fold increase in wild type versus 1.05 in *zip*-RNAi), and a 65% reduction in total myosin (Fig 4G) (32) We assessed the mechanical state of the *zip*-RNAi embryos which revealed a distinct breakdown in the relationship between the cell shape index and cell shape alignment compared to controls *(*Fig. 2G). *zip*-RNAi cells adopted fluid-predictive shapes and alignment but exhibited solid-like behavior with minimal rearrangements (cells falling above the predicted curve, Fig. 4H). These results indicate that *zip*-RNAi embryos show severely reduced tissue fluidity that is not reflected in the cell shape values. Taken together, *Kr* and *zip*-RNAi disrupted GBE with decreased tissue dynamics, providing favorable conditions to examine how tissue mechanical properties influence neuroblast ingression.

**Extended model recapitulates tissue dynamics of experimental phenotypes**

The standard shape-based anisotropic vertex model fails to explain the experimental phenotypes we observed, particularly the decoupling of cell shape from rearrangement dynamics seen in *zip*-RNAi embryos. To address this limitation and create a framework for investigating NB ingression mechanics, we tested whether our vertex model with T1 cellular rearrangement delays could reproduce the two distinct mechanical tissue elongation rate and rearrangement rate defects observed in *Kr* and *zip*-RNAi embryos.

For control simulations, parameters were chosen to reproduce wild-type tissue dynamics (Methods). To simulate *Kr* mutants, we reduced the anisotropic line tension strength by 50% compared to controls, reflecting the mild disruption in myosin levels (Fig. 3F,G). This parameter adjustment resulted in moderate reductions in both tissue elongation rate (Fig. 5A-B) and rearrangement rate (Fig. 5C). For *zip*-RNAi simulations, we implemented two parameter changes: increasing the T1 cellular rearrangement delay by 10-fold (Methods) to capture the severe rearrangement defects and reducing anisotropic line tension strength by 75% to reflect the severe myosin depletion (Fig. 4F,G). These adjustments successfully reproduced the complete loss of tissue elongation and near-absence of T1 transitions observed experimentally (Fig. 5A-C).

With these calibrated parameters reproducing two qualitatively distinct tissue mechanical states, we incorporated NB ingression into the simulations to test how altered tissue mechanics affect NB apical constriction. We modeled NBs undergoing progressive area reduction (Methods, linear tension dynamics, Movie S1). The model predicted substantial ingression delays only in the *zip*-RNAi knockdowns, with the *Kr* mutant simulations having almost identical NB ingression time compared to controls (Fig. 5D-E, Movie S4). In *zip*-RNAi simulations, near-complete tissue solidification predicted slower ingression (Fig. 5D-E, Movie S4). Thus, our model forecasts that near-complete tissue solidification increases mechanical resistance to NB shape changes, as neighboring cells in solid-like tissues cannot rearrange to accommodate apical constriction.

**NB ingression proceeds normally despite reduced tissue fluidity**

To assess the impact of tissue mechanical changes on NB ingression, we quantified individual NB ingression dynamics in control, *Kr* mutant, and *zip*-RNAi embryos (Fig. 6A,C). We limited our analysis to NBs that start ingressing during the first 20 min of GBE, when tissue-level effects were most pronounced. We tracked ingressing NBs within the ventral column, aligned them developmentally to the start of GBE, and selected only NBs that initiated ingression during this period.

To compare ingression kinetics, we aligned all NBs to the end of ingression (apical cell area 2.5 $\mu m^2$) and tracked cell size for 25 minutes prior to the end of ingression (Fig 6B,D). Contrary to our model predictions, we found no significant changes in overall NB ingression speed in either *Kr* or *zip*-RNAi conditions (Fig. 6B',D'). We separated NB ingression into "early" (10-25 min prior to completion) and "late" (final 10 min) phases, the latter corresponding to the period when NBs rapidly remove their apical domains and lose apical-basal polarity (43). *Kr* mutants showed a modest but significant delay during the late phase of ingression (Fig. 6B''). In contrast, *zip*-RNAi showed no significant delays in either phase (Fig. 6D''). Thus, *Kr* mutants, which caused only moderate tissue-level defects, affected NB ingression timing, while *zip*-RNAi, which caused severe tissue solidification, did not.

**Modelling indicates that cell-intrinsic mechanisms support NB ingression dynamics**

The independence of NB ingression from tissue fluidity was unexpected given our model predictions and the dramatic disruption of tissue fluidity through *Kr* and *zip* perturbations. This discrepancy suggests that NBs possess cell-intrinsic mechanisms that maintain robust ingression despite tissue-level mechanical variability. We first assessed whether the highly anisotropic contractility of individual NB contributes to the robustness of ingression dynamics and then examined the role of endocytosis of the apical domain in regulating ingression speed.

NB ingression is normally anisotropic, with AP/vertical junctions shrinking earlier and faster than DV/horizontal junctions (35). Since *Kr* mutants specifically disrupt the planar polarization of myosin (22), we modified our vertex model to test whether loss of NB-intrinsic myosin anisotropy could explain the *Kr* ingression delay by comparing simulations with anisotropic versus isotropic edge tension around NBs undergoing apical constriction. Loss of NB-intrinsic anisotropic tension significantly delayed NB ingression (Fig. 7A, B, Movie S4). NBs with isotropic edge tension ingressed slower than those with anisotropic tension in our simulations, consistent with the delay in late ingression observed in *Kr* mutants (Fig. 6A,B).

Moreover, we confirmed experimentally that both *Kr* and *zip*-RNAi exhibit loss of anisotropic junction shrinkage, with edges distributed across a broader range of angles compared to controls, which maintained edges oriented primarily along DV (0–30°) or AP (60–90°) axes (Fig. S2, MovieS5). This effect was more pronounced in *zip*-RNAi than in *Kr* mutants. Applying isotropic junctional contractility in the *zip*-RNAi model delays NB ingression time further (Fig. 7C,D, Movie S6). Yet experimentally, *zip*-RNAi NBs constrict at normal speed despite both the loss of myosin and anisotropy. This discrepancy suggests an alternative mechanism to ensure robustness of ingression dynamics.

Endocytosis of the apical domain plays a key role in NB ingression (33). In contrast to the severe depletion of myosin, a reduction of endocytosis is currently the only known manipulation that can prevent NB ingression (32, 33). Thus, endocytosis operates at least largely independently of myosin activity in promoting ingression. To test whether endocytosis could explain normal ingression speed under conditions of reduced myosin, we incorporated a coupled endocytosis-contractility model for ingressing NBs, where endocytosis enhances constriction efficiency by contributing to effective tension accumulation (Methods). This model successfully captured the experimentally observed normal ingression behaviour in *zip*-RNAi embryos (Fig. 7C,D, Movie S6). Our model predicts that the effective increase in apical tension reflects not an increase in myosin levels, which are reduced by *zip*-RNAi, but rather the progressive endocytic removal of apical and junctional membrane components such as Crumbs and E-cadherin (32, 33), which reduces resistance to constriction and effectively increases the rate of apical constriction, a mechanism supported by prior experimental work (43). This suggests that modulating the cell-intrinsic endocytic dynamics enables NBs to maintain normal ingression speed even when total myosin levels are severely reduced, anisotropic contractility is abolished, or tissue-level fluidity is compromised.

**Discussion**

Our integrated computational and experimental approach suggests that NB ingression and germ band extension operate as mechanically independent modules despite their spatial and temporal overlap. This independence challenges initial predictions from vertex modeling, which suggested that tissue-level mechanical properties should constrain individual cell behaviors. Instead, cell-intrinsic mechanisms within NBs determine ingression kinetics independent of tissue-level mechanics. Strikingly, robustness of ingression kinetics is maintained even under conditions of near-complete tissue solidification. Our findings suggest that cell-intrinsic endocytosis can override dramatic variations in the tissue mechanical environment.

The characteristic anisotropic distribution of myosin, with enrichment along anterior-posterior edges, generates differential tension that drives the "letterbox" morphology and directional apical constriction of NBs observed in control embryos (35). Loss of this anisotropy in *Kr* mutants disrupts the coordinated edge dynamics that normally accelerates ingression, specifically during late-phase constriction, when the apical domain of NBs is removed. Also, *zip*-RNAi embryos tended to have reduced constriction speed during late NB constriction, although differences from control embryos were not significant. Notably, *Kr* is a pair-rule patterning gene (40), in contrast to *zip,* which encodes a structural component of the myosin motor (22, 44). We speculate that the effects of Kr loss on NB ingression reflect a disruption of planar polarity of myosin within NBs, leading to isotropic constriction and ingression delay, as captured by our isotropic tension model.

In *zip*-RNAi embryos, despite severe myosin depletion and tissue solidification, which was predicted to substantially delay ingression, NBs maintain near-normal apical constriction rates. Our modeling suggests that this behavior could be explained through endocytotic removal and thus reduction in size of the apical domain of NBs. Progressive endocytic removal of apical membrane components may also reduce resistance to constriction, allowing the same myosin activity to produce greater deformation (43). Collectively, our findings predict that NB-specific cell-intrinsic regulation of endocytosis could contribute to maintaining normal ingression speed even under conditions of near-complete tissue solidification. The molecular basis for this relationship between endocytic regulation and ingression speed remains an important question for future investigation.

While tissue-level forces influence cellular behaviors in other developmental contexts (16, 25, 45), the single-cell, localized nature of NB ingression contrasts sharply with most studied examples of mechanical coupling, which typically involve interactions between two large tissue domains with shared interfaces, such as mesoderm-ectoderm interactions (25) or tissue budding as observed in salivary gland (46) and formation of the intestine (47) in *Drosophila*. The mechanical modularity observed with NB ingression may represent a general design principle in developmental systems, enabling robust execution of essential morphogenetic events despite the mechanical complexity and variability inherent in morphogenesis.

Our experiments challenge shape-based theories of the solid-to-fluid transition (11). The previous framework predicts tissue fluidity based on cell shape index and alignment during GBE (11). However, reduction of myosin activity caused tissue solidification while exhibiting fluid-predictive cell geometries. In contrast, *Kr* mutants showed solid-predictive geometries while maintaining some fluidity. This decoupling of cell shape and fluidity in both *zip*-RNAi and *Kr* mutants reveals that cell geometry alone cannot predict mechanical tissue state. Incorporating cellular rearrangement time enabled our models to recapitulate tissue-level dynamics across these perturbations.

We implement NB ingression through progressive area reduction, which does not capture three-dimensional morphology or the detailed dynamics of apical constriction and basal delamination. Additionally, we treat myosin-driven tension as quasi-static and do not explicitly model the periodic pulsatile contractions observed experimentally. While time-averaged contractile forces appear sufficient to capture overall ingression behavior, whether pulsatility contributes mechanistic advantages remains untested. Future models incorporating three-dimensional deformation, adhesion dynamics, and explicit temporal myosin pulse dynamics may better explain ingression robustness.

Key questions remain. How is cell-intrinsic myosin anisotropy established and maintained during active tissue deformation? Does the surrounding epithelium instruct NB planar polarity, or does specification trigger cell-autonomous polarization? While our results suggest mechanical independence, we cannot exclude mechanochemical feedback in which tissue mechanical state influences NB contractile dynamics. Whether endocytic activity in NBs is upregulated in myosin compromised embryos to compensate for reduced actomyosin-based contractility remains an important prediction to test. Alternatively, endocytosis may be the key driver of NB apical domain reduction (apical constriction) given that compromised endocytosis, in contrast to loss of myosin activity, can prevent NB ingression (32,33). Additionally, our study focuses on the first wave of NB ingression. Whether mechanical independence persists during later waves, when the germband is fully extended and which occur therefore in different mechanical contexts, remains unclear. Analysis across developmental stages may reveal how cellular mechanisms adapt to varying tissue-level environments.

Our findings suggest that NB ingression operates largely independently of tissue-level mechanics during germband extension, with cell-intrinsic myosin anisotropy and endocytic dynamics predicted by our model to contribute to ingression robustness under conditions of severe myosin depletion. Spatial and temporal overlap between morphogenetic processes does not imply mechanical coupling, suggesting principles that may broadly govern developmental robustness. Understanding the determinants of mechanical coupling versus independence across developmental contexts remains an important open question.

## Materials and Methods

### Vertex model with convergent extension and ingressing cells

Our anisotropic vertex model with cellular rearrangement time and ingressing cells extends the standard vertex model (38) to include anisotropic line tensions, T1 transition delays (36) and NB cell ingression. The total energy of the tissue follows

$$E = \sum_{\alpha=1}^{N}\left[K_A(A_\alpha - A_{0\alpha})^2 + K_p(P_\alpha - P_{0\alpha})^2\right] + \sum_{ij}\Gamma_{ij}\, l_{ij} \quad (1)$$

where $K_A$ and $K_P$ are area and perimeter elasticity constants, $A_\alpha$ and $P_\alpha$ are cell area and perimeter, $A_{0\alpha}$ and $P_{0\alpha}$ are preferred cell area and preferred cell perimeter. The area term represents incompressibility of the cell and the perimeter term represents the competition between adhesion and contractility. The final term sums over all cell edges, where $l_{ij}$ is the edge length between cells i and j, and $\Gamma_{ij}$ represents the line tension applied to that edge. The total line tension $\Gamma_{ij}$ combines two distinct contributions:

$$\Gamma_{ij} = \gamma_{ij}^{CE} + \delta_{ij}\Gamma_{ij}^{NB} \quad (2)$$

where $\gamma_{ij}^{CE}$ is the convergent-extension tension and $\Gamma_{ij}^{NB}$ is the tension that applies only to neuroblast edges. $\delta_{ij}$ equals one if edge $ij$ belongs to a neuroblast cell, otherwise $\delta_{ij}$ is zero. The anisotropic line tension $\gamma_{ij}^{CE}$ models planar-polarized myosin distribution during convergent extension according to

$$\gamma_{ij}^{CE} = \gamma_0 \cos[2\,(\theta_{ij} - \theta_{AP})] \quad (3)$$

where $\gamma_0$ controls myosin anisotropy strength, $\theta_{ij}$ represents the orientation of the edge, and $\theta_{AP}$ defines the anterior-posterior axis orientation. This tension pattern favors shrinkage of junctions oriented perpendicular to the AP axis, promoting convergent extension through coordinated T1 transitions that drive tissue elongation.

For neuroblast ingression dynamics, we implement a line tension $\Gamma_{ij}^{NB}$ that applies only to edges belonging to neuroblast cells and evolves over time after a defined onset time. This is consistent with the observation that myosin levels in NBs are far in excess of neighboring non-ingressing cells (35), motivating a cell-specific tension term distinct from the tissue-level convergent extension tension. The neuroblast-specific tension can be applied in two distinct modes to capture different biological mechanisms. In isotropic constriction mode, tension increases uniformly on all neuroblast cell edges to model uniform cortical contraction. In anisotropic constriction mode, tension enhancement is preferentially applied along edges aligned with the AP axis to represent directional actomyosin-based constriction observed *in vivo*.

**Neuroblast tension dynamics**

**Linear tension dynamics.** Neuroblast line tension evolves continuously according to

$$\Gamma_{ij}^{NB}(t) = \gamma_{ij}^{NB}(t) + \zeta_{ij}(t) \tag{4}$$

where the deterministic neuroblast tension component increases linearly as

$$\frac{d\gamma_{ij}^{NB}}{dt} = \Delta\gamma^{NB} \tag{5}$$

where $\Delta\gamma^{NB}$ is the tension accumulation rate and remains constant throughout ingression. $\gamma_{ij}^{NB}(0)$ is the initial tension on neuroblast edges at the onset of ingression. Stochastic noise $\zeta_{ij}$ is added at each time step to $\gamma_{ij}^{NB}$, drawn from a Gaussian distribution with 2% standard deviation relative to $\gamma_{ij}^{NB}$, reflecting biological variability in contractile activity. This linear scheme applies to all simulations except *zipper* knockdown simulations incorporating endocytosis-contractility coupling (Fig. 7D).

**Endocytosis-contractility coupling dynamics.** Based on the mutual dependency between actomyosin contractions and endocytosis demonstrated by Simões et al. (43), we modeled their interaction as a coupled system where apical contractions promote endocytic removal of membrane components, and endocytosis in turn reduces resistance to constriction and increases the rate of effective tension accumulation. Tension relaxes between contractile pulses and endocytic activity undergoes natural turnover. For this coupling model, the total NB tension is:

$$\Gamma_{ij}^{NB}(t) = \gamma_{ij}^{NB,eff}(t) + \zeta_{ij}(t)$$

$$\frac{d\gamma_{ij}^{NB,eff}}{dt} = \Delta\gamma^{NB} + \alpha\phi_E - \kappa\gamma_{ij}^{NB,eff} \tag{6}$$

$$\frac{d\phi_E}{dt} = \beta\gamma_{ij}^{NB,eff} - \lambda\phi_E$$

where $\gamma_{ij}^{NB,eff}$ represents effective NB edge tension capturing both myosin-driven contractility and endocytic resistance removal, $\phi_E$ represents endocytic activity, $\alpha$ the endocytic contribution to tension accumulation, $\beta$ the contraction-driven endocytosis rate, $\kappa$ the tension relaxation rate, and $\lambda$ the endocytic turnover rate.

When a neuroblast cell reaches 10% of its initial area through progressive constriction, it is considered fully ingressed and is removed from the tissue.

**Overdamped dynamics.** The vertex dynamics are governed by overdamped equations of motion. The position of each vertex $\boldsymbol{r}_k$ is updated according to

$$\Delta\boldsymbol{r}_k = \mu\boldsymbol{F}_k\Delta t + \boldsymbol{\eta}_k \tag{7}$$

where $\boldsymbol{F}_k = -\nabla_k E$ is the force on vertex $k$, $\mu$ is the inverse friction coefficient, $\Delta t$ is the time step, and $\boldsymbol{\eta}_k$ is a normally distributed random force with zero mean and variance $2\mu T\Delta t$. The effective temperature $T$ introduces Brownian fluctuations in vertex positions (48). Length is nondimensionalized by the natural unit by $l = \sqrt{A_0}$. We define a characteristic timescale $\tau = 1/(\mu K_A A_0)$, and choose an integration step $\Delta t = 0.01\tau$, corresponding to 100 timesteps per characteristic time.

We implement T1 transition delays by imposing a finite rearrangement time $t_{T1}$ that acts as a kinetic brake on cellular neighbor exchanges (36). When local cell dynamics shrink an edge length below a critical threshold for T1 initiation, the affected edge is prevented from completing the transition for the duration $t_{T1}$.

**Simulation implementation and parameters.** Our implementation was built upon the open-source framework cellGPU (49), extended to include tissue elongation, cellular rearrangement delays and ingressing cells. All simulations were initialized with 256 cells in a square periodic box. Initial cell configurations were generated from random Voronoi tessellations and equilibrated for $10^3$ $\tau$ before applying condition-specific parameters. Five cells were randomly selected from the central region of each simulation to represent ingressing NBs.

For quantitative analysis and comparison across conditions, we measured NB ingression time as the duration required for cell area to lose 80% of its initial value. All results represent averages over 40 independent simulations, with NB-specific measurements averaged across 5 NBs per simulation.

For simulations incorporating convergent extension dynamics, the target shape index began at $p_0 = 3.65$ and increased gradually with $\delta p = 0.00001$ per timestep to model the shape index increase observed experimentally. Each simulation continued until the simulation box extended 4-fold along one dimension or until 200 $\tau$ were reached, whichever occurred first.

To examine the effect of T1 cellular rearrangement time (delay) on NB ingression time (Fig. 1A-D) during tissue extension, we initialized simulations with an anisotropic line tension $\gamma_0 = 0.2$. Neuroblast tension evolution followed the linear tension dynamics. The NB-specific line tension began at $\gamma_{ij}^{NB}(0) = 0.1$ and increased with $\Delta\gamma^{NB} = 0.06$ per $\tau$.

For isotropic tissue simulations (Fig. 1E), the T1 transitions were instantaneous as in the standard vertex models, while the target cell shape index $p_0$ (cellular perimeter divided by square root of area) was systematically varied across simulations. Anisotropic edge tension was eliminated by setting $\gamma_0 = 0$.

To reproduce the tissue dynamics observed in control and mutant embryos, we calibrated model parameters based on experimental measurements. Wild-type (control) simulations used a T1 cellular rearrangement time (delay) of 9.5 $\tau$ and anisotropic line tension $\gamma_0 = 0.2$, representing baseline wild-type parameters. For *Kr* mutant simulations, we set T1 delay to $9.5\ \tau$ but reduced the anisotropic line tension to $\gamma_0 = 0.1$, corresponding to 50% of control levels to reflect the disrupted myosin observed experimentally (Fig. 3G). For *zip*-RNAi simulations, we dramatically increased the T1 delay to $103\ \tau$ to capture the severe rearrangement defects and reduced the anisotropic line tension to $\gamma_0 = 0.05$, corresponding to 25% of control levels to reflect severe myosin depletion (Fig. 4G). Other parameters are $\gamma_{ij}^{NB}(0) = 0.1$ and $\Delta\gamma^{NB} = 0.06$ per $\tau$.

For *Kr* mutant isotropic simulations (Fig. 7A,B blue), we use isotropic NB constriction, reflecting the observed loss of myosin anisotropy.

For the endocytosis-contractility coupling model (Fig. 7C,D green), we similarly used isotropic NB constriction, reflecting the loss of myosin anisotropy observed in *zip*-RNAi conditions. The coupled model parameters: $\gamma_{ij}^{NB,eff}(0) = 0.1$, $\Delta\gamma^{NB} = 0.06$, $\phi_E(0) = 0.0$, $\alpha = 0.4$, $\beta = 0.03$, $\kappa = 0.045$ and $\lambda = 0.1$.

Across all simulations, other parameters in simulation units: area modulus $K_A = 1$, perimeter modulus $K_P = 1$, motility coefficient $\mu = 1$, effective temperature $T = 0.02$, and timestep $\Delta t = 0.01\tau$.

**Markers and mutants**

The following fly markers and mutants were used: *ubi-ph::mCherry(50)*, *yw; endoD-Ecad::GFP* (51), *yw; ;* [KrIf-1]/*CyO* (BDSC 92615), *sqh-Gap43::mCherry* (52), *sqh-Sqh::gfp (53)*, *matatub67-Gal4 ;matatub15-Gal4* (a gift from D. St. Johnston, University of Cambridge, Cambridge, England UK), *UAS-zip-RNAi* (Trip line HMS01618; *Drosophila* RNAi Screening Center), Spider::GFP (a gift from J. Zallen, Memorial Sloan Ketterring Cancer Center, New York, NY). To analyze *Kr* mutants we analyzed the F2 progeny *yw; Kr/CyO* females crossed to *ubi-ph::mCherry* or *sqh-Gap43::mCherry, sqh-Sqh::gfp* males. *Kr* mutants were identified phenotypically by GBE defects, with a mix of heterozygote *Kr* mutants and homozygous wildtype embryos being used as control. To knock down myosin heavy chain, we crossed matαtub*67-Gal4 ;matatub15-Gal4* / *matatub67-Gal4 ;matatub15-Gal4*,sqh-sqh::mch females crossed to *UAS-zip-RNAi females* for cell tracking or myosin quantification respectively. These progenies were outcrossed to either *Spider::GFP* or *yw; endoD-Ecad::GFP* and their progeny analyzed. For control, we crossed maternal Gal4 females with Spider::GFP or *yw; endoD-Ecad::GFP*.

**Time-lapse imaging**

Embryos were dechorinated for 3 min in 9.6% hypochlorite solution, transferred to a drop of halocarbon oil 27 on a coverslip and mounted on an oxygen-permeable membrane. Cell tracking experiment images were acquired using a confocal Leica Sp8 microscope with a HCX Plan-Apochromat 63x/1.4 NA CS2 objective. GFP was excited with an optically pumped semiconductor laser (488nm; 4%), and mCherry was excited with an optically pumped semiconductor laser (514nm; 3-5%). 12-bit images of one- or two-color z stacks (15 planes) were acquired at 0.5-um steps in 15- or 10-s intervals using Application Suite X software (Leica Microsystems). For myosin experiments, images were acquired using a confocal Zeiss LSM780 microscope using a Plan-Apochromat 63x/1.4NA M27 objective. GFP was excited (488nm; 8%), and mCherry was excited (514nm; 5%). 12-bit images of one- or two-colour z stacks (20 planes) were acquired at 0.5-um steps in 30- or 40-s intervals using Zen software (Zeiss). Images were maximally projected for analysis in ImageJ (54). Pixel dimensions range from 140 to 190 nm/pixel.

**Tissue Elongation Measurement**

Tissue elongation was measured by tracking 10 different cells per embryo during germband extension. Posterior tissue flow was calculated as the distance the cell traveled relative to its starting position at time 0 for each frame. An average of these 10 cells is presented per embryo.

**Automated Image Segmentation and Tissue Morphology Quantification**

Projected time-lapse movies were smoothed with a gaussian blur prior to segmentation. Processed movies were segmented and analyzed using SEGGA, and errors were corrected manually (55). Cells were tracked and analyzed between t = -2.5 and t = 20 min for each movie. To be included in the cell rearrangement analysis, cells must be in the region of interest for at least 5 minutes after t= 0. The cell rearrangement rate is shown as a uniformly weighted average over 2 minutes. We computed the average cell shape index ($p_0$) at each time point by quantifying both the perimeter P and area A of each segmented cell. Mean cell shape index is the average of P/sqrt(A) over all segmented cells.

Cell shape alignment (Q) was calculated using the triangle method (11). Briefly, cell centers are connected by a triangular network, and the shape stretch of each triangle is quantified. Q is a tensor that captures both the degree of cell elongation and the alignment of elongated cells across the tissue. Q= 0 indicates randomly oriented cells with no net alignment, while higher values indicate cells that are both elongated and aligned in a common direction. Specifically, Q is computed from the average deformation of triangles connecting cell centers, such that tissues with the same degree of cell alignment but more elongated cells have a higher Q.

**Neuroblast Segmentation, Tracking and Quantification**

We used Pyjamas to automatically identify cell outlines in time-lapse videos with a watershed algorithm as described previously (56). To segment a single NB, we manually assigned one seed per NB and all its surrounding neighbors at the earliest time point that the cell of interest could be tracked (stage 7 or early stage 8) using the Set fiducial tool. Seeds were propagated and expanded between consecutive time points, resulting in cell segmentation. Seeds were manually corrected as needed. To measure mean apical surface area during ingression NBs were temporally aligned based on the time when they reached an apical area of 2.5 $\mu m^2$. Onset of ingression was determined manually as the time point at which the initial mean area of the NB started to decline persistently. We approximated NB ingression speed by measuring the delta in apical surface area between the initial mean area and 2.5 $\mu m^2$ divided by the time.

To quantify individual cell edges we manually traced cell interfaces using Pyjamas for each edge that is maintained until the end of ingression. To quantify junctional myosin protein levels, we utilized a 3-pixel-wide dilation of the cell outline and quantified the pixel intensity of the myosin channel under the line. Protein levels were normalized by subtracting the fluorescence mode for the entire image to remove background.

**Statistics**

For the experiments, mean values were determined based on the n value of embryos/cells/cell edges as indicated in each figure legend. Box plots show the interquartile range (IQR, Q1–Q3) as boxes, with the mean indicated by a line; whiskers extend to 1.5×IQR. We used R statistical packages to perform Mann-Whitney U tests to determine p-values.

For computational modeling, shaded regions indicate standard error of the mean (s.e.m.) from n = 40 independent simulations. Box plots show the interquartile range (IQR, Q1–Q3) as boxes, with the mean indicated by a line; whiskers extend to 1.5×IQR. NB-specific measurements were averaged across 5 NBs per simulation. To determine p-values, we performed Mann-Whitney U tests.

**Acknowledgments**

We are grateful to Sergio Simoes for critical reading of the manuscript. We thank the Imaging facility of the Department of Cell and Systems Biology, University of Toronto for support. This study was supported by funds from the Canadian Institutes for Health Research (to U.T., G.E.-T. and R.F.-G.), the Natural Sciences and Engineering Research Council of Canada (to G.E.-T.), and the Strategic Support for Tri-Council Success Seed Grant of Western University (to G.E.-T.). U.T. is a Canada Research Chair for Epithelial Polarity and Development. This research has been enabled in part by the use of computing resources provided by Digital Research Alliance of Canada.

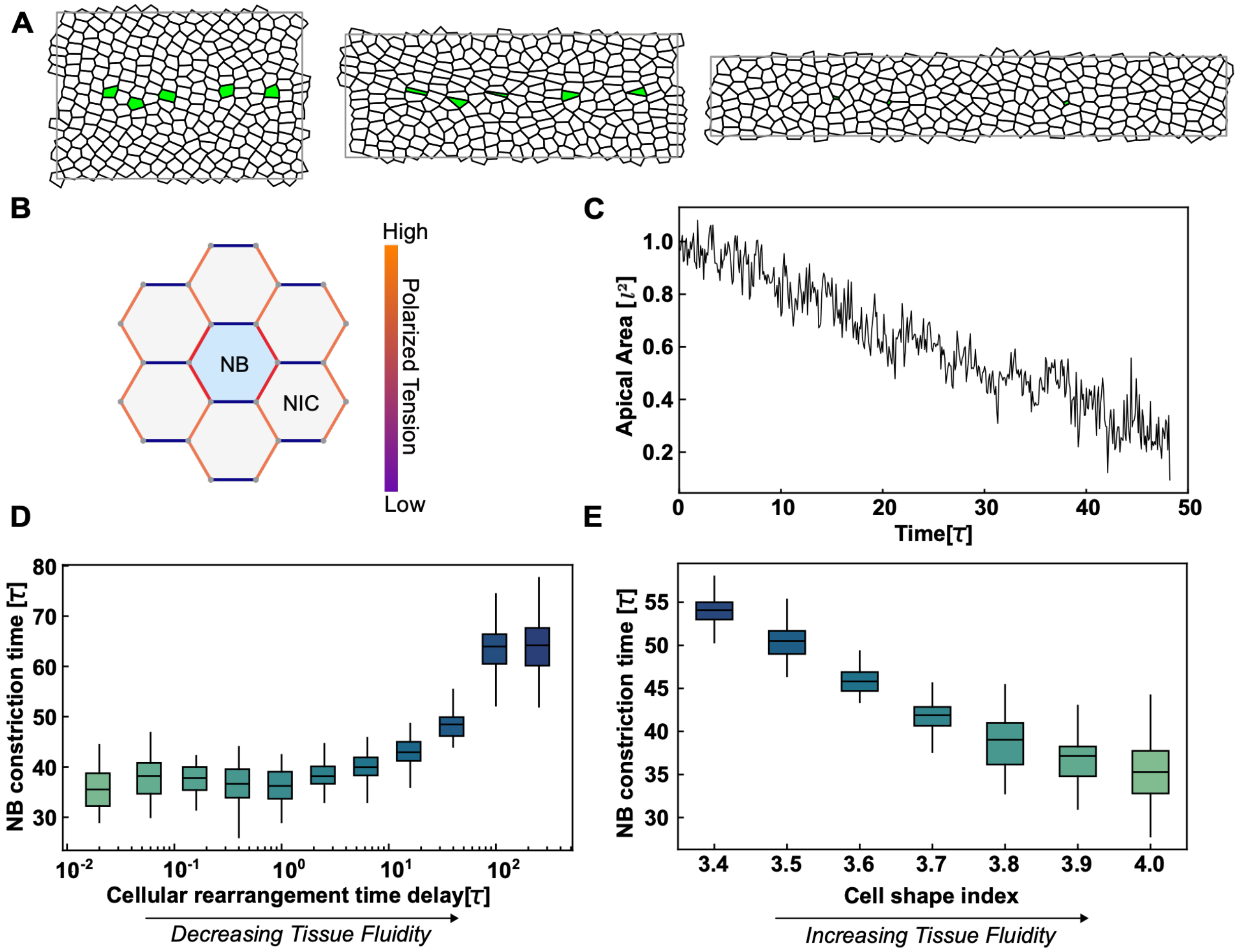

**Figure 1. Reduced tissue fluidity slows NB ingression in the model.** (A) Representation of anisotropic vertex model undergoing convergent extension and NB ingression simultaneously (Movie S1). The polarized tension drives the tissue extension. Cells with coloured face (blue) are the randomly selected NB cells in the middle strip of the tissue that experience apical constriction. (B) Diagram of an NB cell surrounded by non-ingressing cells (NIC). Orange edges indicate junctions with higher polarized tension along AP edges. Red edges of NB indicate higher anisotropic myosin levels of NB (Methods). (C) Area of a single NB cell during tissue elongation with initial time cell shape index $p_0 = 3.65$, $\gamma_0 = 0.2$, and $\gamma_{ij}^{NB}(t = 0) = 0.1$ and T1 delay of 9.53 $\tau$ (see Methods for parameter descriptions). (D) Dependence of NB ingression time on tissue fluidity, controlled by varying T1 delay (cellular rearrangement time) parameter for initial time cell shape index $p_0 = 3.65$, $\gamma_0 = 0.2$, and $\gamma_{ij}^{NB}(t = 0) = 0.1$ (see Methods for parameter descriptions). The ingression times are averaged over 5 cells in each simulation and over 40 independent simulations. (E) Dependence of NB ingression time on tissue fluidity, controlled by cell shape index ($p_0$) alone for isotropic simulations (see Methods for other parameters). The ingression times are averaged over 5 cells in each simulation and over 40 independent simulations. Box plots: IQR (Q1–Q3); line = mean; whiskers = 1.5×IQR.

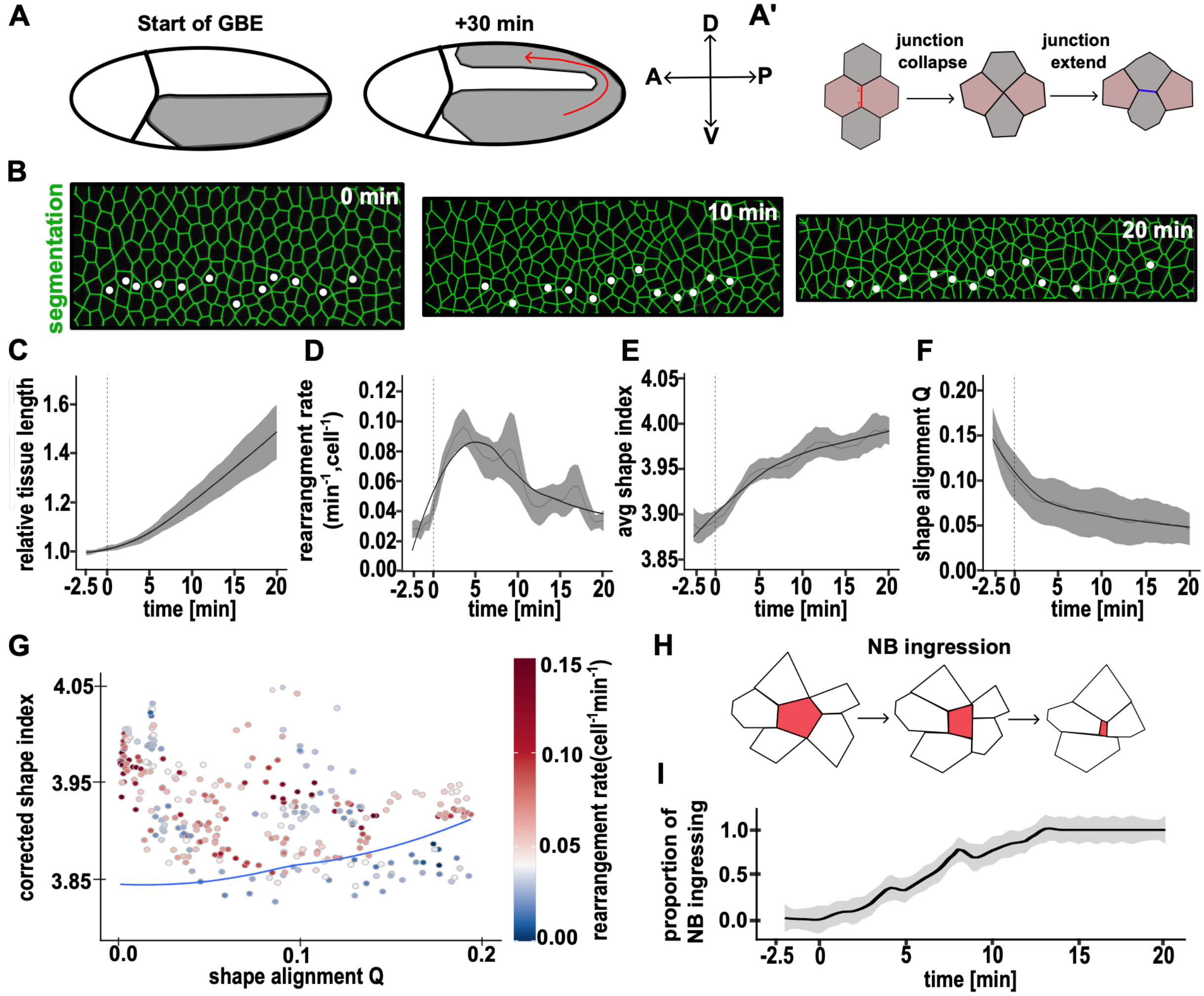

**Figure 2. NB ingress in a fluid-like environment.** (A) Schematic of *Drosophila* body axis elongation. The germband (grey) elongates along the head-to-tail body axis. (A') Schematic of oriented rearrangement that occurs during GBE. (B) Stills from time-lapse video of the ventral epithelium (t=0, start of GBE). (membrane marker PH::mCherry). Anterior left, ventral down. Images were overlaid with polygonal segmentations used to quantify cell shape and dynamics (green) using SEGGA. (C-F) Quantification of tissue dynamics versus time aligned to the start of GBE. Dark line indicates a loess-fitted line, faint line represents the mean with a 2-minute rolling average, shaded region indicates the sem. n = 7 wild-type embryos with an average number of 105 cells measured per embryo per time point. (B) Tissue elongation, (C) instantaneous rearrangement rate, (D) average cell shape index, (E) tissue anisotropy. (G) The relationship between corrected cell shape index (11) and cell shape alignment; each point represents a single time point in a single embryo. Instantaneous cell rearrangement rate is represented by color per point. The solid line indicates the parameter-free prediction of transition which was previously tested in Wang 2020 (11). n = 7 embryos. (H) Schematic of NB apical constriction. (I) Quantification of proportion of ventral column NBs that have started ingression from 2 thoracic segments. n= 18 NBs.

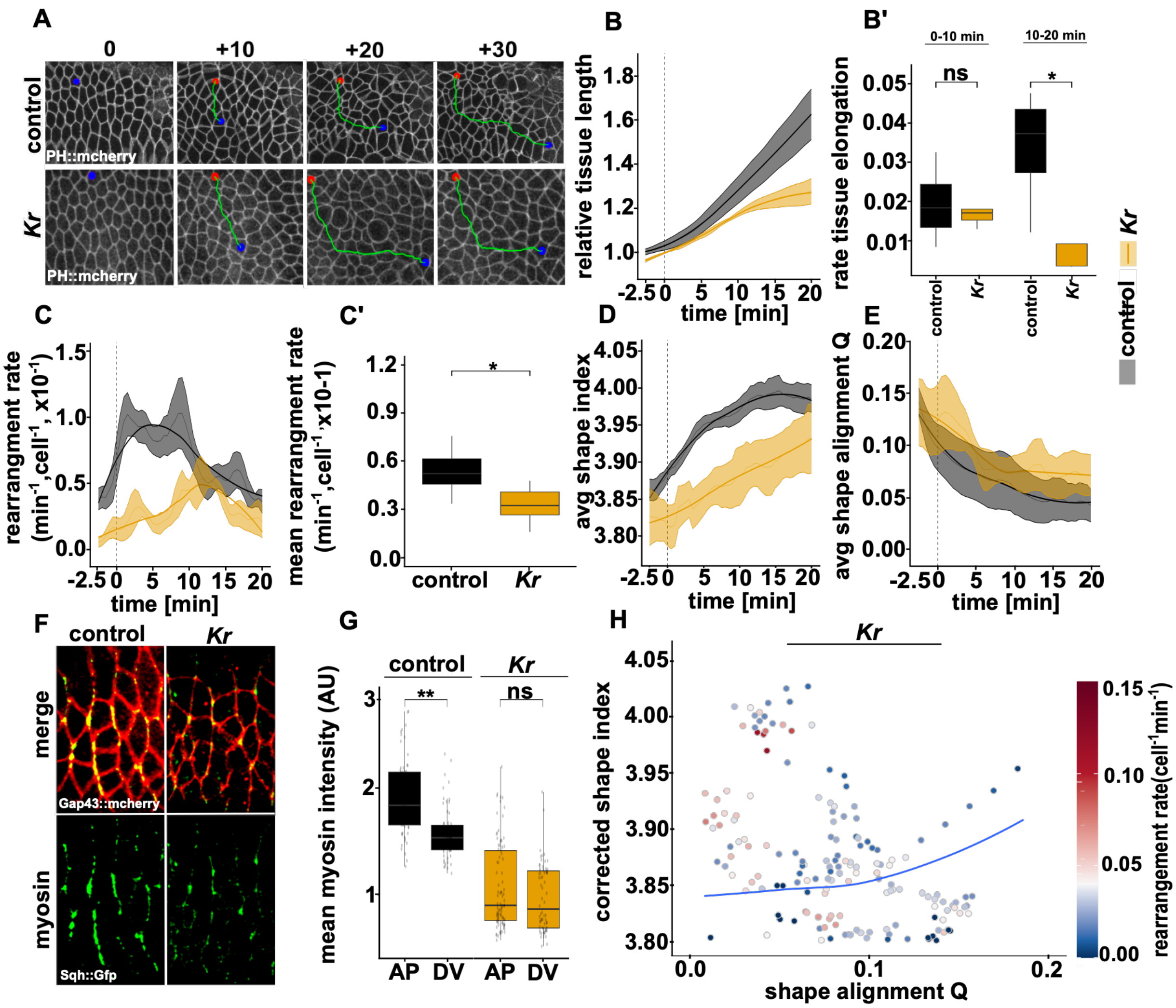

**Figure 3.** ***Kr* mutants moderately reduce tissue fluidity during GBE.** (A) Representative stills from live imaging of ventral epithelium of stage 7-9 embryos (membrane marker *PH::mCherry)*. Green line shows migration of a single cell at the onset of GBE (0min, red point) to 30 minutes later (blue point). (B-E) Quantification of tissue dynamics over time, aligned to the start of GBE in *Kr* mutants compared to wild-type (from Figure 2). Dark line indicates a loess-fitted line; faint line represents the mean with a 2-minute rolling average; shaded region indicates the s.e.m. n = 4 for controls (Kr/+, +/+ embryos) and 5 for Kr mutant embryos, with an average of 114 cells measured per embryo per time point. (B) Relative tissue length over time. (B') Rate of tissue elongation during the first 20 minutes of GBE. (C) Instantaneous rearrangement rate over time. (C') Mean rearrangement rate during the first 20 minutes of GBE. (D) Average shape index over time. Box plots with IQRs and minimum/maximums displayed. (E) Average cell shape alignment (tissue anisotropy) over time. (F) Confocal images of ventral epithelium at stage 8 embryos in *Kr* mutant and wild-type embryos showing myosin levels (Sqh::Gfp) and membrane markers (GAP43::mCherry). (G) Quantification of myosin intensity of individual edges from data in F. Edges were binned into AP (70-90) or DV(0-20°). Each data point represents an individual edge. (H) The relationship between corrected cell shape index (11) and cell shape alignment; each point represents a single time point in a single embryo. Instantaneous cell rearrangement rate is represented by colour per point. The solid line indicates the parameter-free prediction of transition which was previously tested in Wang 2020 (11). p-values: ns = no significant difference, * < 0.05, ** < 0.001, Mann-Whitney U test. Box plots: IQR (Q1–Q3); line = mean; whiskers = 1.5×IQR.

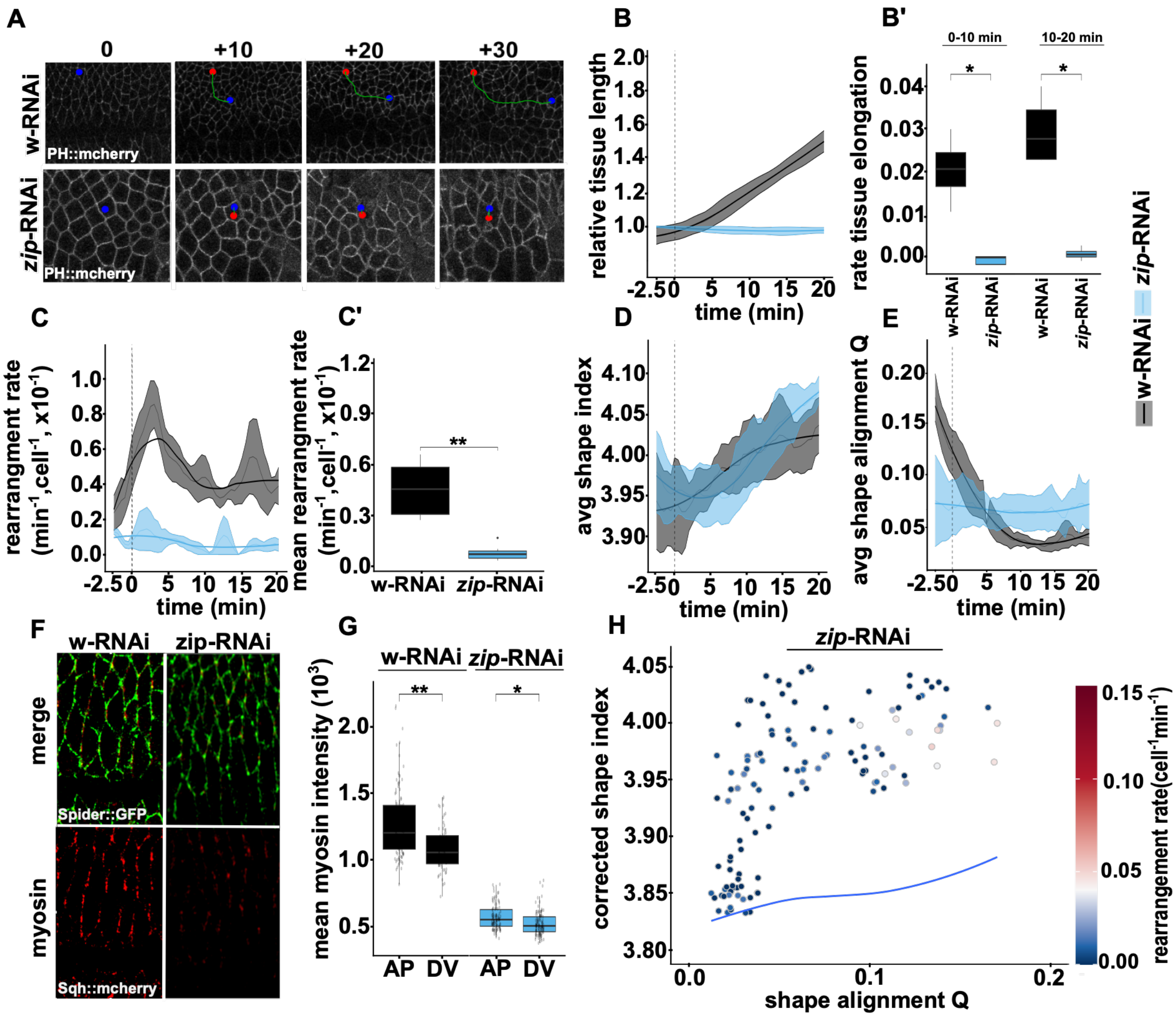

**Figure 4. *zip*-RNAi shows near-complete tissue solidification during GBE.** (A) Representative stills from live imaging of ventral epithelium of stage 7-9 embryos (membrane marker Spider::GFP*)*. Green line shows migration of a single cell at the onset of GBE (0min, red point) to 30 minutes later (blue point). (B-E) Quantification of tissue dynamics over time, aligned to the start of GBE in zip-RNAi compared to wild-type (from Figure 2). Dark line indicates a loess-fitted line; faint line represents the mean with a 2-minute rolling average; shaded region indicates the s.e.m. n = 4 embryos for control (w-RNAi) and 4 embryos for zip-RNAi, with an average of 95 cells measured per embryo per time point. (B) Relative tissue length over time. (B') Rate of tissue elongation during the first 20 minutes of GBE. (C) Instantaneous rearrangement rate over time. (C') Mean rearrangement rate during the first 20 minutes of GBE. Box plots with IQRs and minimum/maximums displayed. (D) Average shape index over time. (E) Average cell shape alignment (tissue anisotropy) over time. (F) Confocal images of ventral epithelium at stage 8 embryos in *zip-RNAi* and w-RNAi embryos showing myosin levels (Sqh::mCherry) and membrane markers (Spider::GFP). (G) Quantification of myosin intensity of individual edges from data in F. Edges were binned into AP (70-90) or DV(0-20°). Each data point represents an individual edge. (H) The relationship between corrected cell shape index (11) and cell shape alignment; each point represents a single time point in a single embryo. Instantaneous cell

rearrangement rate is represented by colour per point. The solid line indicates the parameter-free prediction of transition which was previously tested in Wang 2020. p-values: ns = no significant difference, * < 0.05, ** < 0.001, Mann-Whitney U test. Box plots: IQR (Q1–Q3); line = mean; whiskers = 1.5×IQR.

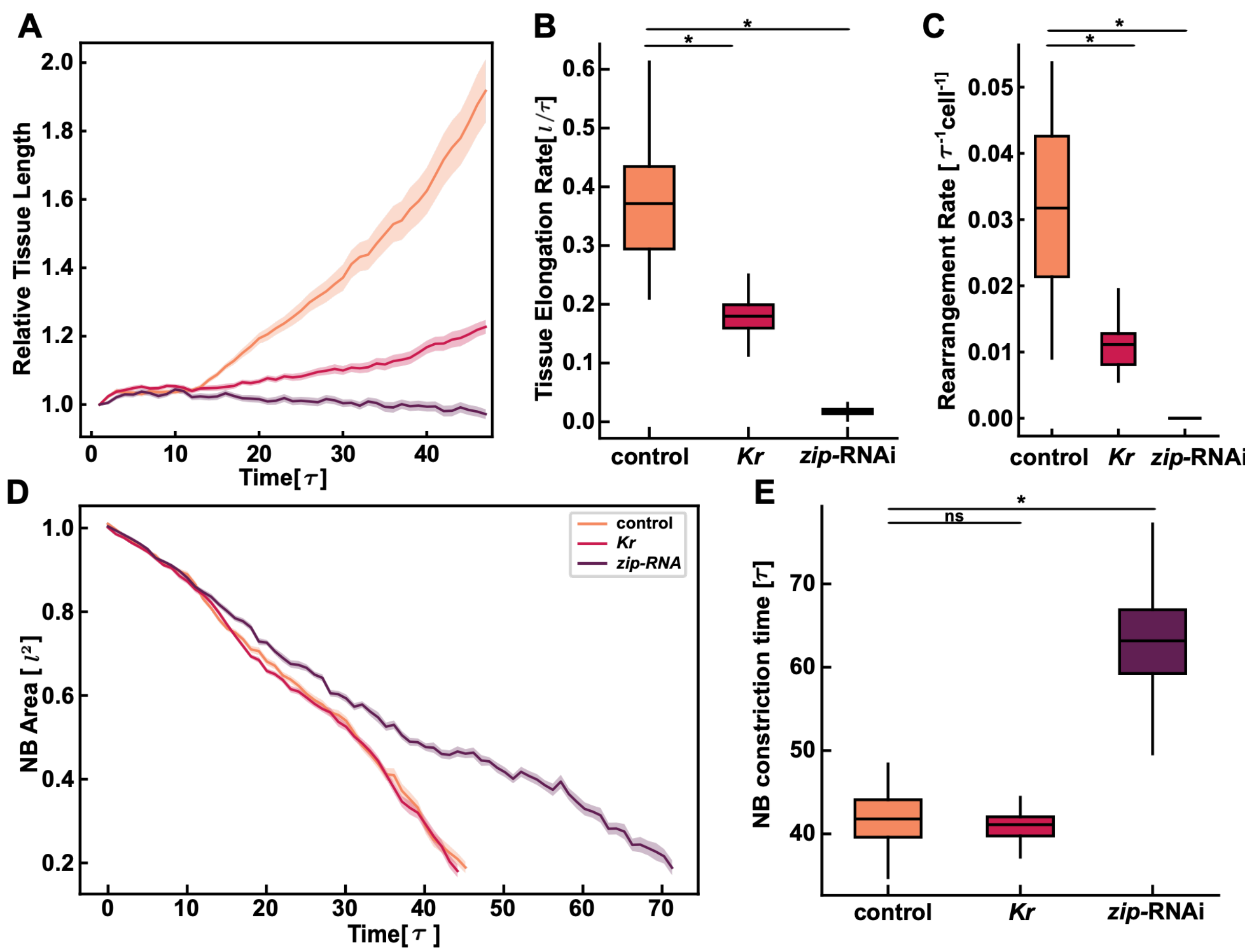

**Figure 5. Extended model with cellular rearrangement delays recapitulates tissue dynamics and predicts ingression defects in zip-RNAi embryos.** (A) Tissue elongation length over time and (B) tissue elongation rate in control (orange), *Kr* mutant (red), and *zip*-RNAi (purple) simulations ($n = 40$ simulations). Control simulations used baseline parameters (*Methods*). *Kr* simulations implemented 50% reduction in anisotropic line tension strength ($\gamma_0$) compared to control, resulting in moderate elongation defects. *zip*-RNAi simulations combined increased T1 transition delay with 75% reduction in anisotropic line tension ($\gamma_0$) compared to control (Methods). (C) T1 rearrangement rate across control and mutant simulations. (D) Representative simulations of neuroblast (NB) ingression showing mean area reduction over time in control, *Kr* mutant, and *zip*-RNAi simulations. Error bars, s.e.m. ($n = 40$ simulations). (E) Quantification of NB constriction time across conditions. The ingression time is averaged over 5 cells in each simulation and over 40 independent simulations. Box plots: IQR (Q1–Q3); line = mean; whiskers = 1.5×IQR. *Mann-Whitney U test, $p < 0.001$, ns = no significant difference.

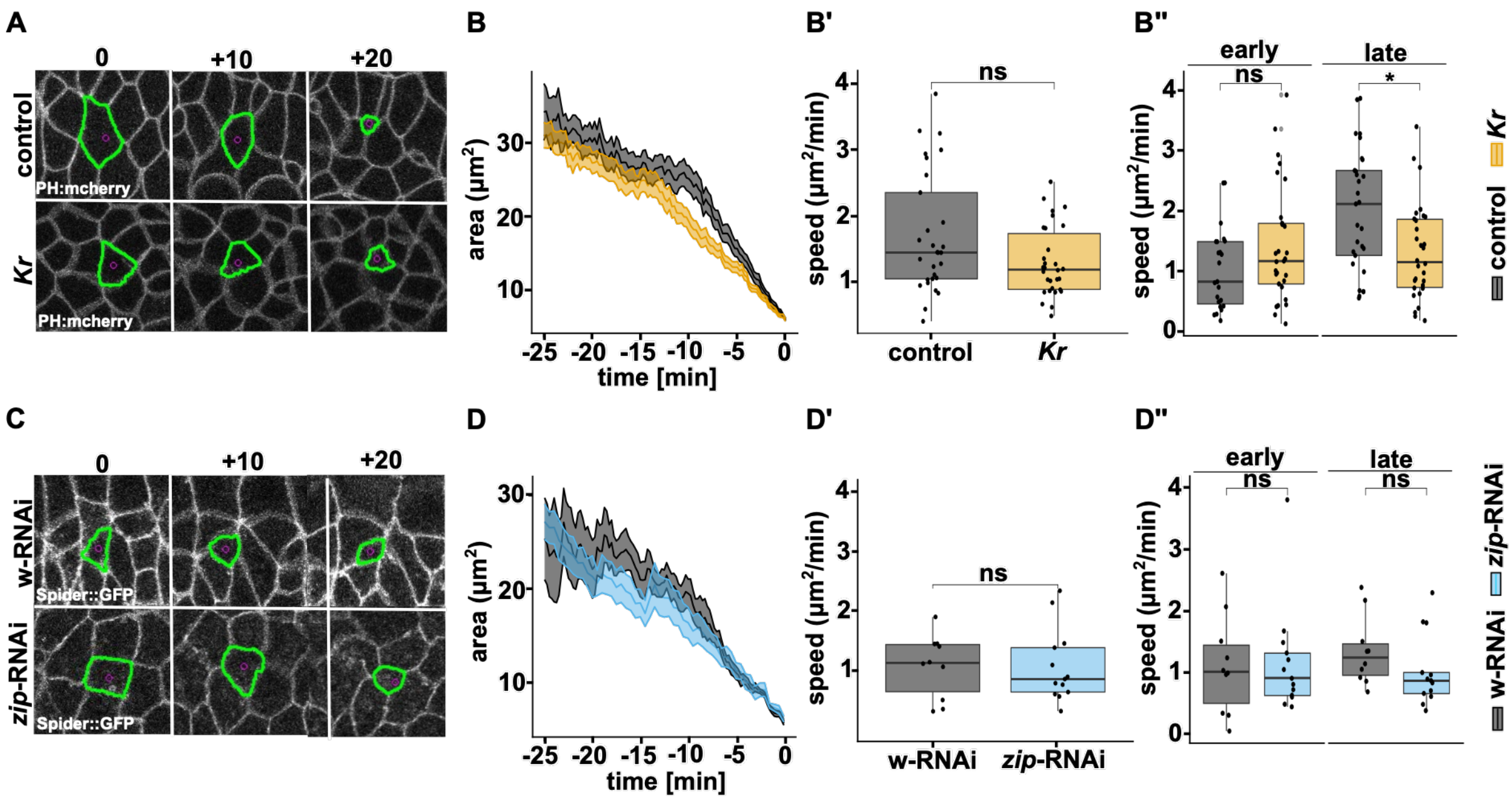

**Figure 6. Delay in Neuroblast ingression does not correlate with tissue fluidity dynamics**. (A, C) Representative time-lapse imaging of NB ingression. Representative stills show neuroblasts (NBs) ingressing from the ventral neuroectoderm at 0, +10, and +20 min. NBs were segmented by watershed-based membrane segmentation and tracked over time. (A) Kr mutant and control embryos expressing PH::mCherry. (C) zip-RNAi and w-RNAi control embryos expressing Spider::GFP. (B–B", D–D") Quantification of NB area loss and ingression speed.For each genotype, NB apical area was measured over time and aligned to the end of ingression. Quantifications were performed from 4 embryos per genotype, with 20–25 NBs traced per genotype. Shaded regions indicate s.e.m.(B, D) Mean NB apical area over time for each genotype. (B', D') Mean loss of NB apical area over time, aligned to ingression completion. (B", D") NB ingression speed, calculated as the rate of apical area loss in µm²/min, during the early phase of ingression, defined as 10–25 min before completion, and the late phase, defined as the final 10 min before completion. Box plots show the interquartile range with the mean indicated. Statistical significance was assessed using a Mann–Whitney U test; ns, not significant; $*p < 0.05$; $**p < 0.001$.

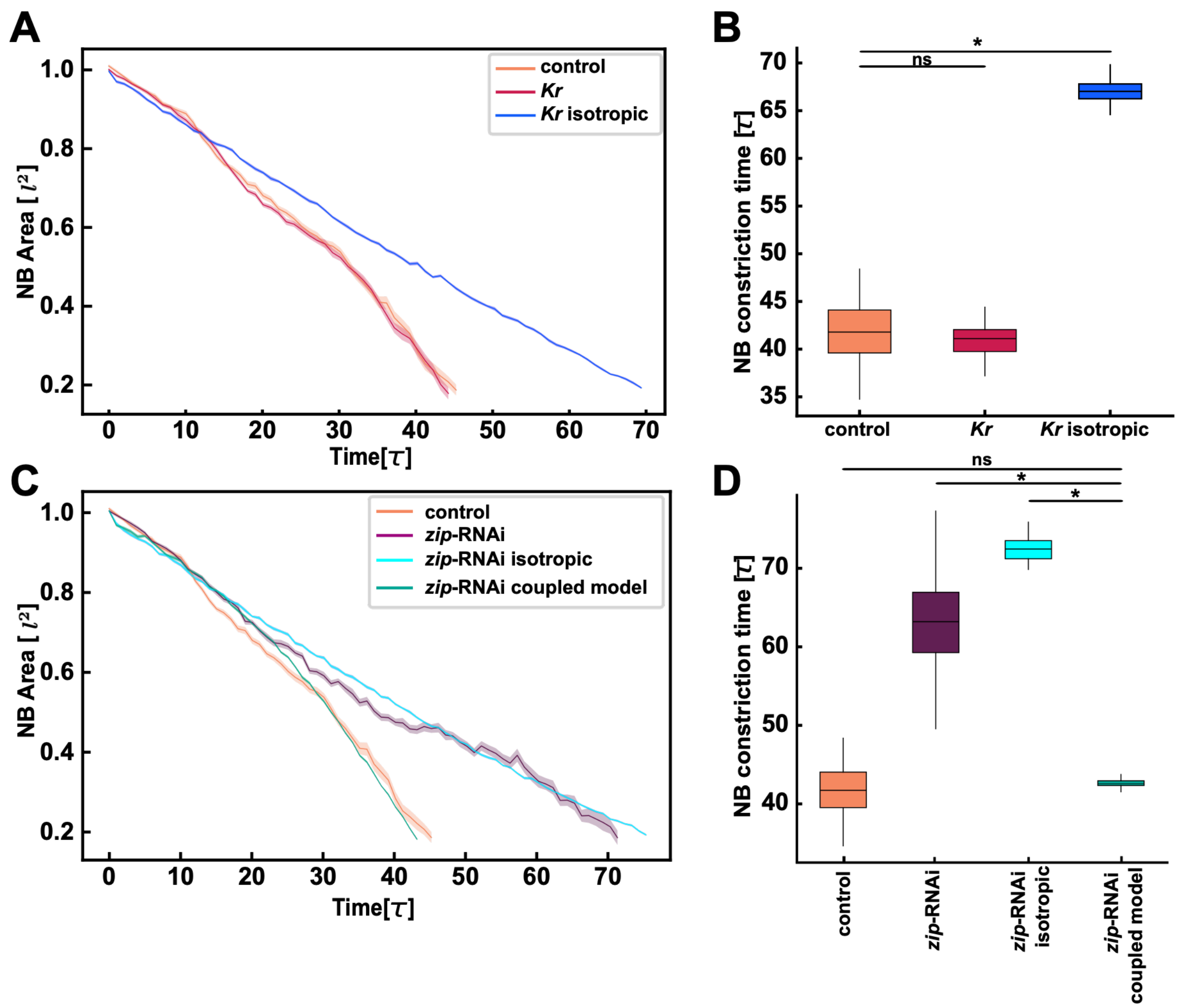

**Figure 7. Cell-intrinsic mechanisms explain NB ingression dynamics in *Kr* and *zip*-RNAi conditions in the model.** (A) NB area over time in simulations with control parameters (orange), *Kr* parameters with anisotropic NB constriction (red) and *Kr* parameters with isotropic NB constriction (blue). (B) Quantification of NB constriction time across conditions. (C) NB area over time in simulations with control parameters (orange), *zip*-RNAi parameters with anisotropic NB constriction (purple), *zip*-RNAi parameters with isotropic NB constriction (cyan) and *zip*-RNAi parameters with endocytosis-contractility coupling (green)**.** (D) Quantification of NB constriction time across conditions. The ingression time is averaged over 5 cells in each simulation and over 40 independent simulations. All simulations except the coupled model used linear tension dynamics (Methods). Control simulations used baseline parameters (Methods). *Kr* simulations implemented a 50% reduction in anisotropic line tension strength ($\gamma_0$) compared to control. zip-RNAi simulations combined increased T1 transition delay with 75% reduction in anisotropic line tension ($\gamma_0$) compared to control, representing more severe myosin depletion. Shaded regions, s.e.m. ($n = 40$ simulations). Box plots: IQR (Q1–Q3); line = mean; whiskers = 1.5×IQR. *Mann-Whitney U test, $p < 0.001$, ns = no significant difference.

## Supplementary Figures

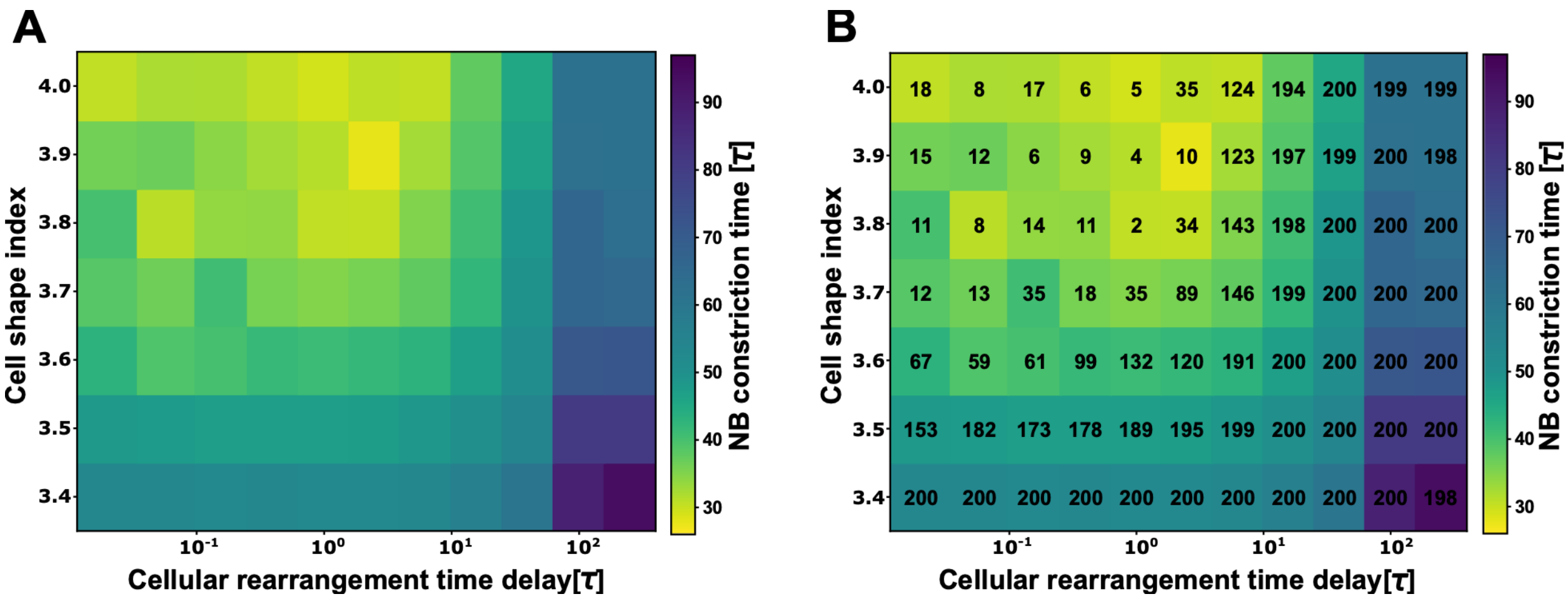

**Figure S1: Combined effects of T1 transition delay and cell shape index on NB ingression timing.** (A) NB constriction time as a function of both T1 cellular rearrangement time (delay) and cell shape index. (B) Same plot as in (A) with the number of successfully ingressed NB cells overlaid in the corresponding parameter grid. Each grid point represents 40 independent simulations with 5 NB cells per simulation (total 200 NB cells per condition). $\gamma_0 = 0.2$, and $\gamma_{ij}^{NB}(t = 0) = 0.1$. Note that missing data occur when either tissue elongation (4-fold) or simulation time limit (200τ) is reached before NB ingression completes, which happens at both extremes of the parameter space: rapid tissue extension (short T1 delay, high shape index) and excessive tissue solidification (long T1 delay, low shape index).

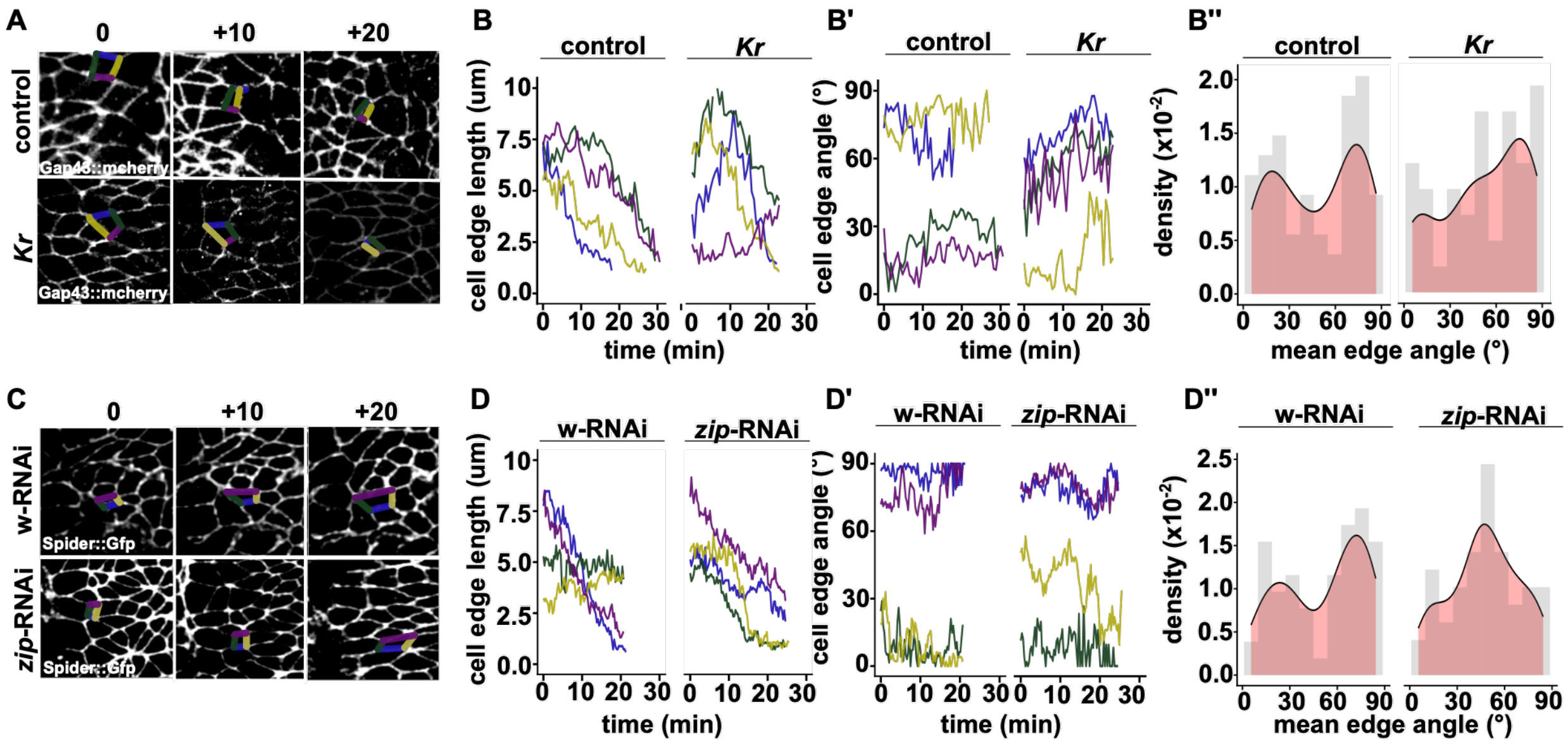

**Figure S2: Loss of anisotropy in *Kr* mutants and *zip*-RNAi.** (A) Representative time-lapse images of ventral epithelium during NB ingression (membrane marker Gap43::mCherry, individual edges color-coded to plots B-B '). Anterior left, ventral down, t= 0 aligned to start of ingression. (B-B'). Quantification of individual edges from NB tracked in panel A. t= 0 aligned to start of ingression. (B) Edge length, (B') cell angle. (B'') Probability density distribution of edge angles. Peaks indicate angles with relative higher frequencies (n = 5 embryos, 30 NBs. (C) Representative time-lapse images of ventral epithelium during NB ingression (membrane marker Spider::Gfp, individual edges color-coded to plots D-D '). Anterior left, ventral down, t= 0 aligned to start of ingression. (D-D') Quantification of individual edges from NB tracked in panel A. t= 0 aligned to start of ingression. (D) Edge length, (D') cell angle. (D'') Probability density distribution of edge angles. Peaks indicate angles with relative higher frequencies n = 4 embryos, 20 NBs per genotype.

**Supplementary Movie Captions**

**Movie S1. Simultaneous convergent extension and NB ingression.** Time-lapse visualization of an anisotropic vertex model showing simultaneous tissue convergent extension (cell intercalation driven by anisotropic myosin organization) and NB ingression (apical area constriction). Five ingressing NB cells are embedded within the tissue (highlighted in green). The tissue undergoes planar-polarized cell rearrangements (T1 transitions) characteristic of convergent extension, progressively elongating 2-fold along the anterior-posterior axis. Simulation parameters correspond to control conditions (Methods).

**Movie S2: Neuroblast ingression occurs concurrently with GBE.** Time lapse of PH: mcherry-expressing embryo tracking the apical surface of individual NBs within the ventral column. Video displays 14 frames per second. Time in minutes:seconds.

**Movie S3: Tissue elongation delays in myosin defective embryos.** Time lapse of PH:mcherry embryos in wild-type, *Kr and zip*-RNAi backgrounds. Green line shows tracking of movement of a single cell during GBE. Time is min:sec aligned to the onset of GBE (t=0). Video displays in 7 frames per second.

**Movie S4. Loss of myosin anisotropy delays ingression.** Time-lapse visualization of the model simulating *Kr* mutant conditions with (right) and without (left) loss of NB myosin anisotropy. Tissue undergoes impaired convergent extension with reduced anisotropic line tension (50% of control). Five ingressing NB cells (green) undergo apical constriction with anisotropic tension along AP edges (left) or with isotropic tension applied uniformly to all edges, reflecting the observed loss of myosin polarization in *Kr* mutants (right).

**Movie S5: Loss of anisotropy in *Kr* mutants and *zip*-RNAi.** Time lapse of embryos in wild-type (Gap43::mCherry), *Kr* (Gap43::mCherry) *and* zip-RNAi (Spider::Gfp) backgrounds. Representative NB is segmented in cyan. Time in min:sec. Video displays 7 frames per second.

**Movie S6. *zip*-RNAi simulations: endocytosis-contractility coupling rescues NB ingression in the model despite tissue solidification.** Time-lapse visualization of the model simulating *zip*-RNAi conditions with near-complete tissue solidification. In all panels, tissue undergoes severe reduction in cell rearrangements (10-fold increase in T1 delay) and dramatically reduced anisotropic line tension (25% of control). Five ingressing NB cells (green) undergo apical constriction under three conditions: anisotropic tension along AP edges (left), isotropic tension applied uniformly to all edges (middle), or isotropic tension modulated by endocytic activity via endocytosis-contractility coupling (right). While anisotropic (left) and isotropic (middle) NB constriction both fail to recover normal ingression timing, the coupled endocytosis-contractility model (right) rescues normal ingression despite severe tissue solidification and myosin depletion.